# Measuring the Variability of Outcomes of the New Natural Daylighting Requirements in the Basque Country's Habitability Decree

**Jorge Otaegi and Iñigo Rodríguez-Vidal**
CAVIAR Research Group, Department of Architecture, University of the Basque Country

*ABSTRACT:* The Basque Country's Habitability Decree of 2022 redefines natural lighting requirements in residential spaces, increasing the Wall-to-Floor Ratio (WFR) compared to previous municipal and sectoral standards, based on the depth of the space considered. This regulatory adjustment seeks to optimize the quality of natural light, a key factor for well-being. However, the application of a geometric rule based on fixed proportions raises questions about the variability of achievable lighting results and their suitability for residential comfort needs. This study analyses the implementation of the regulation in two bedroom models, BR1 and BR2, selected as the typical and the deepest, but with less than 4 meters and meeting the rest of the dimensional requirements, respectively. Through simulations of the Daylight Factor (DF) using window proportions of 1:1, 3:2, 2:1, and 1:2 under a CIE overcast sky model, the dispersion in the obtained natural lighting has been examined. Such analysis allows for the quantification of lighting result variability within the regulatory parameters, exploring the complexity of meeting lighting requirements through an approach based exclusively on geometric proportions. Additionally, the impact of architectural and urban variables such as building height, the presence of courtyards, street width, and balcony depth has been considered. These factors, often overlooked in regulations and indirectly considered in D80/2022, play an important role in modulating natural light and lighting quality. At the end, the paper comments on the advantages and limitations of regulating natural light through arithmetic or geometric rules like the WFR, comparing this method with approaches based on detailed room-by-room analysis.

## INTRODUCTION

Daylight is an essential component of architectural design, contributing to both the visual comfort of building occupants and the overall energy performance of a building. Natural lighting strategies are increasingly prioritized in building codes and sustainability frameworks, as daylight can significantly reduce energy consumption by minimizing the need for artificial lighting and improve indoor environment quality (IEQ). In residential buildings, particularly multi-family housing, the placement and design of windows play a critical role in determining the amount of daylight that enters the space, and thus in the architectural quality and qualities of a dwelling. Achieving the appropriate balance between daylight and other design considerations such as privacy, heat gain and the possibilities for furnishing is a complex process, especially in small apartments and situations where cost is a limiting factor.

In many European regulations, daylighting requirements have remained unchanged for decades. In the Basque Country, the D80/2022 Decree establishes requirements for window-to-wall ratios (WWR) and other parameters related to daylight provision. However, the decree leaves room for variability in several factors such as window proportions, room layout, and external obstructions, which can lead to different daylight outcomes in practice.

This study explores the variability of daylight results under the framework of D80/2022 by performing detailed daylight factor (DF) calculations in bedrooms of multi-family housing. By focusing on common room configurations and window designs, the research aims to provide insights into how different variables affect the overall daylighting performance of these spaces, and how the current regulation does not account for that effect. The study highlights the influence of geometric parameters, such as room depth and window dimensions, as well as physical properties like glass transmittance and external obstructions. By comparing the daylight results across different scenarios, this paper seeks to inform architects and policymakers about the potential impacts of the current regulations and suggest further refinement.

**New regulation in the Basque Country, Spain**

In the Basque Country, the new Decree 80/2022 was born with the vocation to unify the minimum habitability standards for residential buildings and other types of accommodation in the Basque Autonomous Community (CAPV). Prior to this decree, there was no comprehensive regional regulation encompassing all aspects of habitability; existing requirements were scattered across sector-specific laws and municipal ordinances. Preparatory work for the decree began in 2019, leading to its approval in June 2022 and mandatory implementation starting on November 30, 2022. The decree sets forth technical specifications for various aspects of housing design. These include minimum requirements for floor area, ceiling height, equipment, natural lighting, and accessibility among others. Its aim is to ensure that dwellings meet modern standards of functionality, safety, and comfort.

Regarding daylighting, in Decree 80/2022 it is established that rooms intended for living, dining, cooking, and sleeping must have exterior openings for natural lighting, excluding only access areas, bathrooms, pantries, and storage rooms. The surface area of these openings, excluding the window frames, must be a percentage of the usable floor area of the entire illuminated space, varying based on the depth of the space and whether the opening faces the exterior or interior courtyards. For example, for spaces with a depth of less than 4 meters and openings to the exterior, a minimum of 10% is required; if the depth is 4 meters or more, the percentage increases to 15%. If the space faces a courtyard, the requirement is increased and additional 5%

**Other regional and national regulations**

In Spain, the 1944 Order is a milestone regarding the regulation of indoor environmental quality and hygiene conditions in buildings (Ministerio de la Gobernación, 1944). Specifically for lighting, this Order established that every habitable room must have a window with a surface area greater than or equal to 1/6 of the floor area, that is, a WFR (Window-to-Floor Ratio) ≥ 16.66%. In reality, the standard determines the need for a window based on ventilation, but since ventilation was only conceived through windows for habitable rooms like bedrooms and living rooms, we can accept it as the required surface area for lighting.

Most subsequent regulations, whether municipal ordinances or regional norms, have tended to require a significantly lower WFR, as low as 8% (Jiménez Renedo, 2020). Only the autonomous community of Galicia continues to require a WFR of 1/6 in some cases (rooms illuminated through others), although it generally requires 1/8 (WFR of 12.5%) for the most typical situations, like Catalonia. One of the highest requirements, according to the study by (Jiménez Renedo, 2020), is in the Valencian Community, where a WFR of at least 18% is required for rooms deeper than 4 meters illuminated through certain types of courtyards. The following table (Table 1) summarizes the requirements for new buildings in the different autonomous communities:

In line with the trend in Spain, Italian regulations are region-specific but generally require that habitable rooms have a window area of at least 1/8 of the floor area, equating to a WFR of 12.5%. This is the case in Lombardy, Emilia-Romagna, Tuscany and Veneto, for instance.

Germany's regulations focus on both daylight provision and energy efficiency. The DIN 5034 standard outlines the requirements for natural lighting in interiors, suggesting that the window area should be at least 10% of the floor area for adequate daylighting. The Dutch Building regulation stipulates that the glazed area in a habitable room must be at least 10% of the floor area. Moreover, the Dutch decree includes criteria for daylight obstruction from adjacent buildings, ensuring that new constructions do not excessively shade neighboring properties. Denmark's building regulations mandate that the daylight factor in habitable rooms must not be less than 2%, ensuring a high quality of natural light indoors.

The British Standard BS 8206-2 provides guidelines on daylighting, recommending a minimum Average Daylight Factor (ADF) of 2% for kitchens, 1.5% for living rooms, and 1% for bedrooms. The ADF is a metric that considers the amount of daylight received inside a room, factoring in window size, glazing transmittance, and room dimensions.

**Performance-based standards and voluntary certification schemes**

Traditional daylighting regulations often employ prescriptive methods, such as fixed window-to-floor ratios (WFR), to ensure a minimum level of natural light in residential spaces. While these ratios provide clear guidelines and are easy to apply, they may not account for the actual daylight performance in different contexts. In contrast, performance-based standards focus on achieving specific daylighting outcomes using measurable metrics like the Daylight Factor (DF), Spatial Daylight Autonomy (sDA), or Useful Daylight Illuminance (UDI). These metrics assess how effectively natural light illuminates interior spaces over time, considering factors such as window placement, glazing properties, room geometry, and obstructions. By evaluating the actual performance, designers can tailor solutions to unique conditions for each project, enhancing daylight quality while potentially reducing energy need for artificial lighting.

Voluntary green building certification schemes, such as LEED or the WELL Building Standard, incorporate performance-based criteria to promote superior daylighting in residential buildings. LEED v4 Residential includes a Daylight credit that rewards projects demonstrating adequate daylighting levels, which is proved by achieving specified thresholds of sDA, ensuring that a significant portion of regularly occupied spaces receives sufficient natural light without excessive glare. Projects must also control for Annual Sunlight Exposure (ASE). Under the Light concept, WELL emphasizes optimizing daylight exposure to support circadian health (combined with electric lighting), visual acuity, and overall well-being. It sets performance targets for daylight access ASE, spatial Daylight Autonomy (sDA), glare control, and views to the exterior, encouraging designs that connect occupants with the natural environment.

It is logical to us that there is a shift towards performance-based national regulations, where the quality and usability of daylight are considered over simple ratios. It is foreseeable that innovations in glazing, and especially design software allow architects and engineers to optimize daylighting more effectively, leading to more nuanced regulations that account for factors like all the variables considered in this study, plus maybe orientation and climate.

## METHODS

This section presents the methodology used to evaluate the variability of results that arises from the application of the new regulations explained in the introduction. To this end, daylight calculations were performed in bedrooms of multi-family housing, as this type of space typically has a single light source and is usually homogeneous in size and shape. The following subsections define the variables considered, specifying which were treated as fixed and dynamic, the case studies and real-world references used for the calculations and control, and finally, the calculation parameters.

### Fixed and dynamic variables

Of all the physical and geometric parameters that determine the amount of light inside a room, the Basque Decree 80/2022 only controls room depth, $r_d$. and thus generally requires a WWR of 10 or 15%. This approach leaves many variables out of the equation, and thus produces a wide variety of outcomes. Table 2 gathers the variables and parameters that have been taken into consideration in this paper.

Among these variables, some were treated as fixed and others as dynamic, depending on the specific calculation in each case. In general, the geometric parameters corresponding to two typical bedrooms were fixed, as explained further in the subsections that follow.

Specifically, the following calculation scenarios were conducted: (I) with the room dimensions fixed as BR1 and BR2, and all other variables also fixed, varying the window dimensions, or in other words, altering the WFR; (II) keeping the room area constant and fixing either the WFR or window size, calculating the daylight factor (DF) for different room depths, $r_d$; (III) calculating DF and other indicators while keeping all parameters constant except for the light transmittance of the glass, *LT*; (IV) calculating by modifying only the depth of a balcony or overhang placed above the window; and (V) calculating with increasing obstruction angles in front of the window, while keeping all other parameters fixed.

**Selection of reference models and informative cases**

First, it is necessary to establish the format, size, and proportion of the windows to be studied. In practice, the minimum window surface area in each case is defined by Decree D80/2022, but nothing is specified in the regulation about their proportions, height, etc., as mentioned in the introduction. To gain an understanding of the most commonly used shapes, references from 15 ongoing or completed projects were taken. The criterion for selecting the projects was simply availability, so it cannot be considered a true selection, and there was no intention of representativeness. However, all the portrayed windows are from recent multi-family housing projects under some form of protection, whether promoted by the Basque Government itself, its public company Visesa, or by municipalities within the Basque Autonomous Community. Figure 1 shows the resulting bedroom window collection.

It was found that in none of the projects did the windows exceed a height of 2.30m (with the top edge higher than 2.30m above floor level). The width range was between 0.90 and 2.00m, and the total window heights ranged from 1.10 to 2.20m. The height of the windowsills fell into three clearly distinct groups: a large group where the windowsills are between 0.00 and 0.20m from the floor, a smaller group where the windowsill is around 0.50m, and finally, a group of windows with sills in the 0.80-1.10m range. Although the sample was not fully representative, there is a noticeable number of vertical windows, indicating a potential architectural preference.

The glazed area (following the D80/2022 mandate to count only glass surface, omitting the window frames) ranged from 0.93 to 2.00m$^2$, the average being 1.39m$^2$. This range shows a significant degree of variability, as the maximum value is more than double the minimum. However, it must be said that the considered window frame thickness varied considerably from project to project. As a comment, when extracting the windows from their original projects and applying them to the reference rooms BR1 and BR2, which we present in the following paragraphs, the WFR ratio varied between 8.1% and 17.4%, with an average around 12%.

Regarding the selection of the reference rooms for the calculations, it was decided to use the minimum bedroom size from D80/2022, which is 11.50 m$^2$ (10.00 m$^2$ of minimum area + 1.50 m$^2$ for the mandatory personal storage space). Based on this size and considering that D80/2022 establishes different WFR requirements depending on the depth of the room, the reference rooms BR1 and BR2 were defined. BR1 has interior dimensions of 2.50 m wide by 4.60 m deep, so D80/2022 requires a WFR of 15% or higher. Meanwhile, BR2 has a depth of less than 4.00 m, with a width of 2.90 m and a depth of 3.97 m. This is the 'limit bedroom' for which the Basque Habitability Decree allows a WFR of only 10%.The resulting rooms can be seen in figure 2, marked by letters (a) and (b).

The height of the rooms was set to 2.70m, which is the maximum free interior height a typical residential building with a floor height of 3.00 m would allow for, in bedrooms (typically, 2.50-2.70m). The height of the work plane for calculation was assumed to be 0.85m, as shown in (c) in fig. 2.

Daylight calculations in this study used the split-flux method to determine the internally reflected component, while raytracing was used to calculate direct sky and externally reflected components. The sky model used was CIE 16, a traditional CIE overcast sky (ISO 15469:2004). This approach was previously validated through comparisons with Radiance and DIVA/DAYSIM models by the authors of the simulation tool (Marsh and Stravoravdis, 2017). A calculation grid with a 12.50cm spacing, that is, a fine calculation net of points separated 12.50cm in both directions, as shown in Figure 2. Interior surface reflectance was set at 0.40, 0.60 and 0.70 for floor, walls and ceiling, respectively. A constant albedo of 0.20 was considered in all calculations. The dimensions of the rooms for calculation have already been described in subsection 2.2, for both BR1 and BR2. For some calculations, room depth, $r_d$, was progressively increased to study its effect. In the calculations where the room proportions were altered and deviated from BR1 and BR2, the interior usable surface and volume were kept constant, at 11.50m$^2$ and 31.05m$^3$, respectively.

Unless specifically stated in a certain calculation, a light transmittance coefficient of 0.70 was used in all calculations. This obviously obviates the calculations made isolating the light transmittance variable, in which transmittances ranging from 0.30 to 0.90 were tested. Generally, the sill height, $h_s$, was set at 1.00m. In calculation procedures where the size of the window was progressively increased, the sill height was fixed at said until the top border of the window could not continue growing (because it would clash with the ceiling), and only then the sill height was progressively reduced. This was done to not hinder further the results for vertically oriented windows.

Regarding obstacles, apart from the shading depth, $s_d$, depicted in figure 2, all kinds of obstructions were considered calculating the Vertical Sky Component, *VSC*, for each of the simulations. The formula for it can be read in figure 3, alongside its correlation with the obstruction angle from the horizontal, $\gamma_b$. *VSC* represents the ratio of the illuminance at a specific point, caused by direct light from an overcast sky, to the illuminance on an unobstructed external surface under the same sky conditions.

The figure (fig. 3) also provides a general idea of the amount of visible sky vault depending on position inside a same building. A perceptive reader will be able to read in the figure the different influence that the upper and lower obstruction angles, $\gamma_b$ and $\gamma_a$ (as defined in fig. 2), have in the amount of light available. In calculations with standard overcast skies like CIE 1 and CIE 16, access to view of the zenith provides more light than lower angle openings.

In all calculation cases, the average daylight factor, *ADF*, was calculated, together with the minimum and maximum *DF* to calculate uniformity, although these three values are not reported in this paper. The definition of the daylight factor, in its origin (as opposed to derived formulas used for calculation) is shown in equation 1:

$$DF = \frac{E_i}{E_o} \cdot 100\ (\%) \qquad (1)$$

Where $E_i$ and $E_o$ are the inside and outside illuminance, respectively, measured in any inside point at the height of the working plane and in an unobstructed horizontal plane under an overcast sky. In the case of ADF, the spacing of the grid used improves accuracy of the calculation over typical professional compliance reports (Littlefair et al., 2022).

In addition to average, minimum and maximum DF, the portion of area above a DF of 1% and above a DF of 2% was calculated. In the case of the CIE 16 overcast sky, this roughly corresponds to 100 and 200 lux, respectively.

## RESULTS

Figure 4 shows the results of calculating the average daylight factor (DF) for the model rooms BR1 and BR2, modifying the window area or, equivalently, the WFR, up to 50%. It also depicts the minimum DF and the negative effect of a 1-meter deep overhang (balcony). The differing performance of windows with different proportions can be observed. It is evident that, roughly, the average DF increases by 1% for every 5 percentage points of WFR, with diminishing returns as the ratio increases. The curve for windows with a 1:2 proportion performs worse proportionally, as they provide more light below the working plane.

When considering the effect of modifying the light transmittance of the glazing (Aguilar-Santana et al., 2020), as shown in Figure 5, it is evident that it also has a significant impact on the amount of daylight, though it may be less relevant than the window’s format or proportions.

For example, in the case of BR1, the difference between a light transmittance of 50% and 80%, such as the difference between some low-emissivity double-glazing and standard double-glazing, results in more than a 1% difference in DF. Figure 6 illustrates the effect of adding an overhang with increasing depth to rooms BR1 and BR2, equipped with windows with a WFR of 15% and 10%, respectively.

The following graphs in Figure 7 translate the geometric requirements of the Habitability Decree into physical magnitudes such as average DF and the area with a DF > 1% (equivalent to ~100 lux). As shown, there is an abrupt jump in the lighting requirements for rooms with depths under 4 meters compared to those exceeding this threshold. Once again, the poor performance of the vertical window is evident, and the difference compared to other window formats increases as the size (area) of the window grows.

## DISCUSSION AND FUTURE WORK

From direct observation of the results, it can be confirmed that there is a significant variation among windows considered 'equal' or 'equivalent' under D80/2022, at least when measuring the quantity/quality of daylight using the Daylight Factor (DF).

In the near future, there are plans to develop a static calculation sheet, as a simple software tool that is easy to use and does not require a significant time investment from designers. This tool would allow for the verification of average DF calculations in rectangular rooms lit from one side. Additionally, there is consideration of producing an updated brief guide with suggestions for improving natural lighting in typical multi-family housing projects.

## CONCLUSIONS

The analysis of the D80/2022 Decree demonstrates significant variability in daylighting performance for windows considered equivalent under the regulation. While increasing the WFR from 10% to 15% improves daylighting in deeper rooms, our results suggest that the rigidity of this jump in the requirement might be too much. Taller windows or ensuring a greater portion of glazing above the working plane could be more effective.

Window transmittance also showed to influence average daylight factor (DF) very significantly, with differences between installing one type of glazing or the other being equivalent to a 5% change in WFR. External elements such as balconies had a notable impact, significantly reducing DF as their depth increases, as expected.

The translation of geometric ratios into daylighting metrics reveals a gap between regulatory intentions and real-world performance. The WFR alone does not account for the complexity of factors influencing daylight, such as window proportions, placement, and external shading. Addressing these elements in future regulations, and potentially introducing performance-based metrics like daylight factor thresholds, would lead to more consistent and effective outcomes for residential daylighting.

From the observation of recent projects, it was found that a considerable number of them employed narrow, vertically oriented windows, where a significant portion of the glazing (40-50%) was positioned below the working plane, thus providing lighting of limited utility. Additionally, it was noted that windows with lintels higher than 2.30 meters were rarely used, even though a greater height could effectively illuminate a larger part of the room, especially in deeper spaces.

If the goal is to ensure that deep rooms or bedrooms receive adequate lighting, as seems to be the intent in D80/2022, it would be more effective to require taller windows or to prevent more than half of the window area from being positioned below a person's waist, rather than mandating an abrupt 5% WFR increase at the 4-meter depth threshold. It should be noted that the Basque Habitability Decree does not allow the glazed portion of a window above 2.20 meters to be counted as useful for illumination purposes.

Additionally, technicalities such as excluding the frame area and counting only the glazed area had limited effectiveness in our review of 15 projects. The designers of these projects used arbitrary frame thicknesses, which were not checked or adjusted according to the actual thickness once installed. In our view, it would be preferable to establish slightly higher requirements, or as high as necessary, applied to the gross area of the rough window opening, as this dimension tends to remain constant over time. This approach may be debatable and should be aligned with the criteria used in other regions and countries.

## ACKNOWLEDGEMENTS

This document is one of the first works of the Lab for Sustainability and Health in the Built Environment and dissemination of the SDGs in the School of Architecture, under the Campus Bizia Lab (CBL) program of the University of the Basque Country UPV-EHU. The work included in this paper and its publication was funded by the Directorate for Sustainability and Social Commitment of the University of the Basque Country UPV-EHU.

## FIGURES AND TABLES

*(in order of appearance in the text)*

| Region | Regulation | WFR | Region | Regulation | WFR |
|---|---|---|---|---|---|
| Asturias | D 73/2018 | 8.0 % | Extremadura | D 10/2019 | 10.0 % |
| Baleares | D 145/1997 | 10.0 % | Galicia | D 29/2010 | 12.5 – 16.0 % |
| Canarias | D 117/2006 | 8.0 % | Navarra | D 142/2004 | 10.0 % |
| Cantabria | D 141/1991 | 8.0 % | Basque C. | D 80/2022 | 10.0 – 15.0 % |
| Cataluña | D 141/2012 | 12.5 % | La Rioja | D 28/2013 | 10.0 % |
| Madrid* | *NN.UU. PGOU* | 12.0 % | Valencia | D 151/2009 | 10.0 – 18.0 % |

*Table 1. Comparison of required WFR in the Spanish regions. Source: authors based on (Jiménez Renedo, 2020).*

| | | | | | |
|---|---|---|---|---|---|
| $\mathbf{r_d}$ | Room depth | [ m ] | $\boldsymbol{\gamma_c}$ | Complementary angle to $\gamma_a$ | [ º ] |
| $\mathbf{r_w}$ | Room width | [ m ] | $\boldsymbol{\theta}$ | Sky vault view angle | [ º ] |
| $\mathbf{r_h}$ | Room height | [ m ] | **LT** | Light transmittance | |
| $\mathbf{s_d}$ | Shading depth | [ m ] | $\mathbf{R_w}$ | Wall reflectance | |
| $\mathbf{w_w}$ | Window width | [ m ] | $\mathbf{R_f}$ | Floor reflectance | |
| $\mathbf{w_h}$ | Window height | [ m ] | $\mathbf{R_c}$ | Ceiling reflectance | |
| $\mathbf{s_h}$ | Sill height | [ m ] | **A** | Albedo | |
| $\boldsymbol{\gamma_a}$ | Upper obstruction angle | [ º ] | **WWR** | Window to Wall Ratio | % |
| $\boldsymbol{\gamma_b}$ | Lower obstruction angle | [ º ] | **WFR** | Window to Facade Ratio | % |

*Table 2. List of intervening variables, physical parameters and derived metrics.*

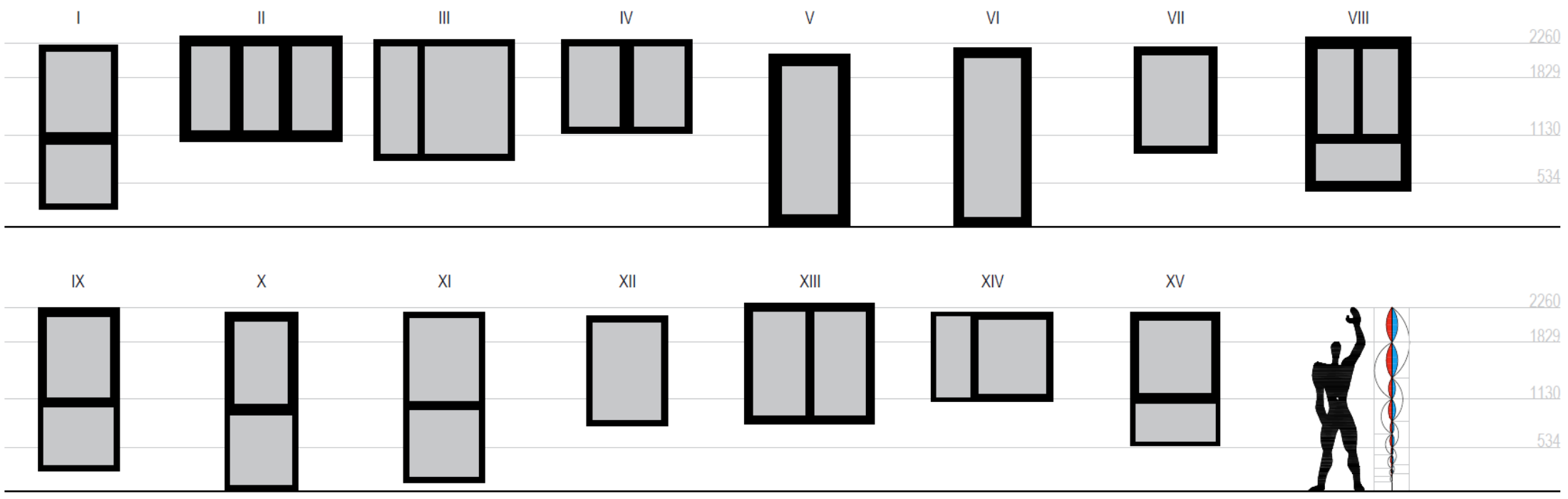

*Figure 1. Collection of windows retrieved from the 15 public housing projects/developments.*

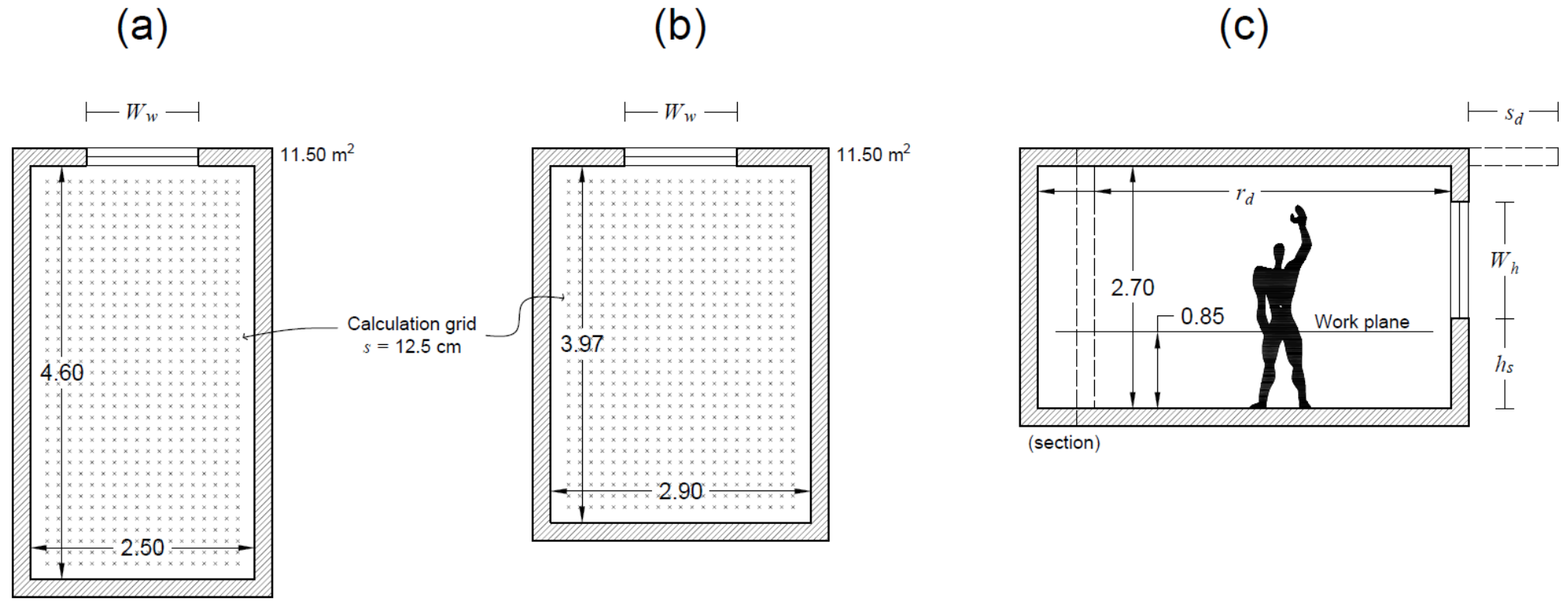

*Figure 2. Bedroom models chosen for calculation and study. BR1 is represented in (a), while BR2 is shown in (b). Picture (c) shows the rooms and parameter designation, in section.*

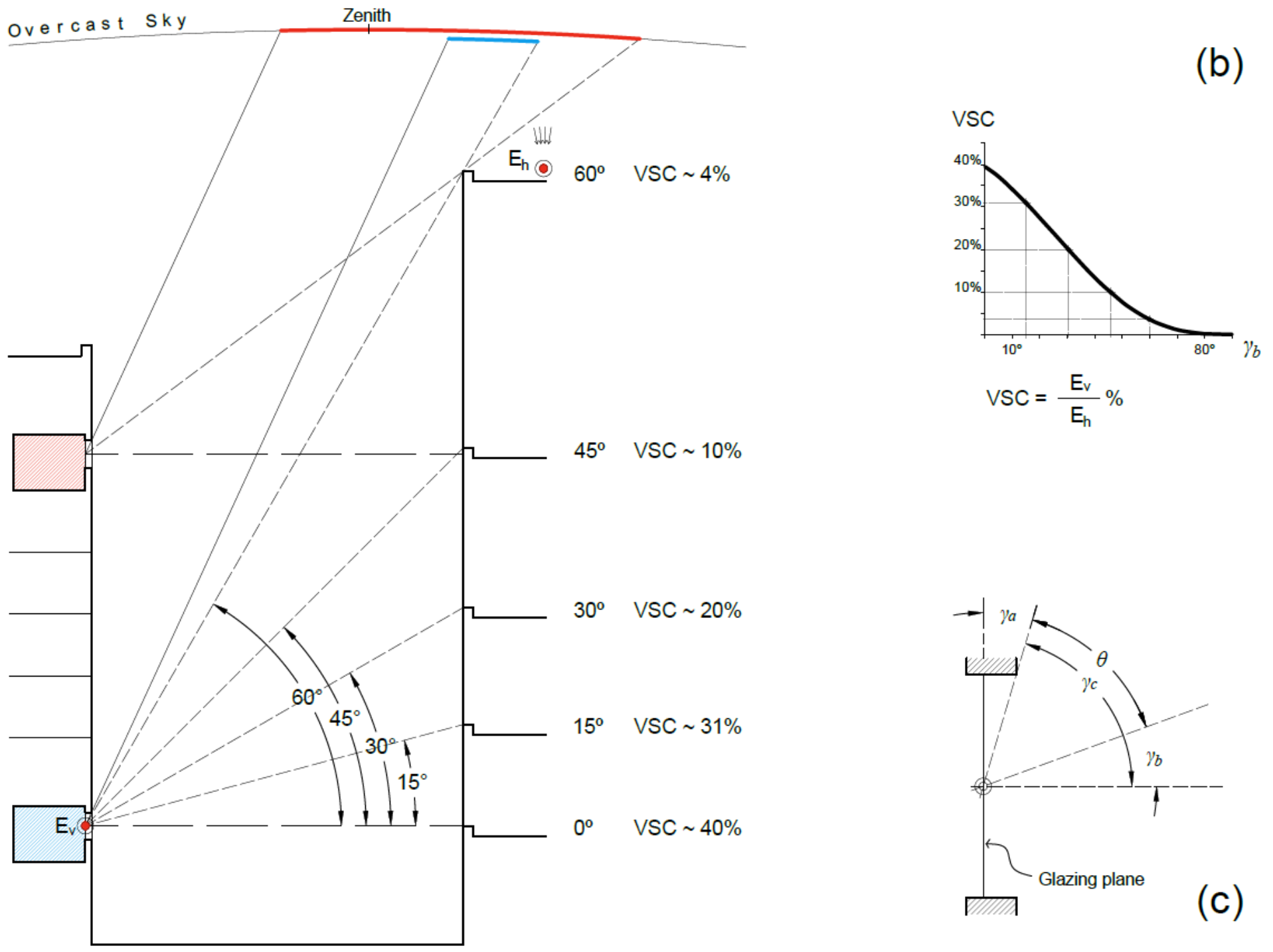

*Figure 3. Picture (a) shows the effect of building, shading or other obstacle obstruction. Pictures (b) and (c) show the typical relation between VSC and the horizontal obstruction angle $\gamma_b$ and the angle convention used, respectively.*

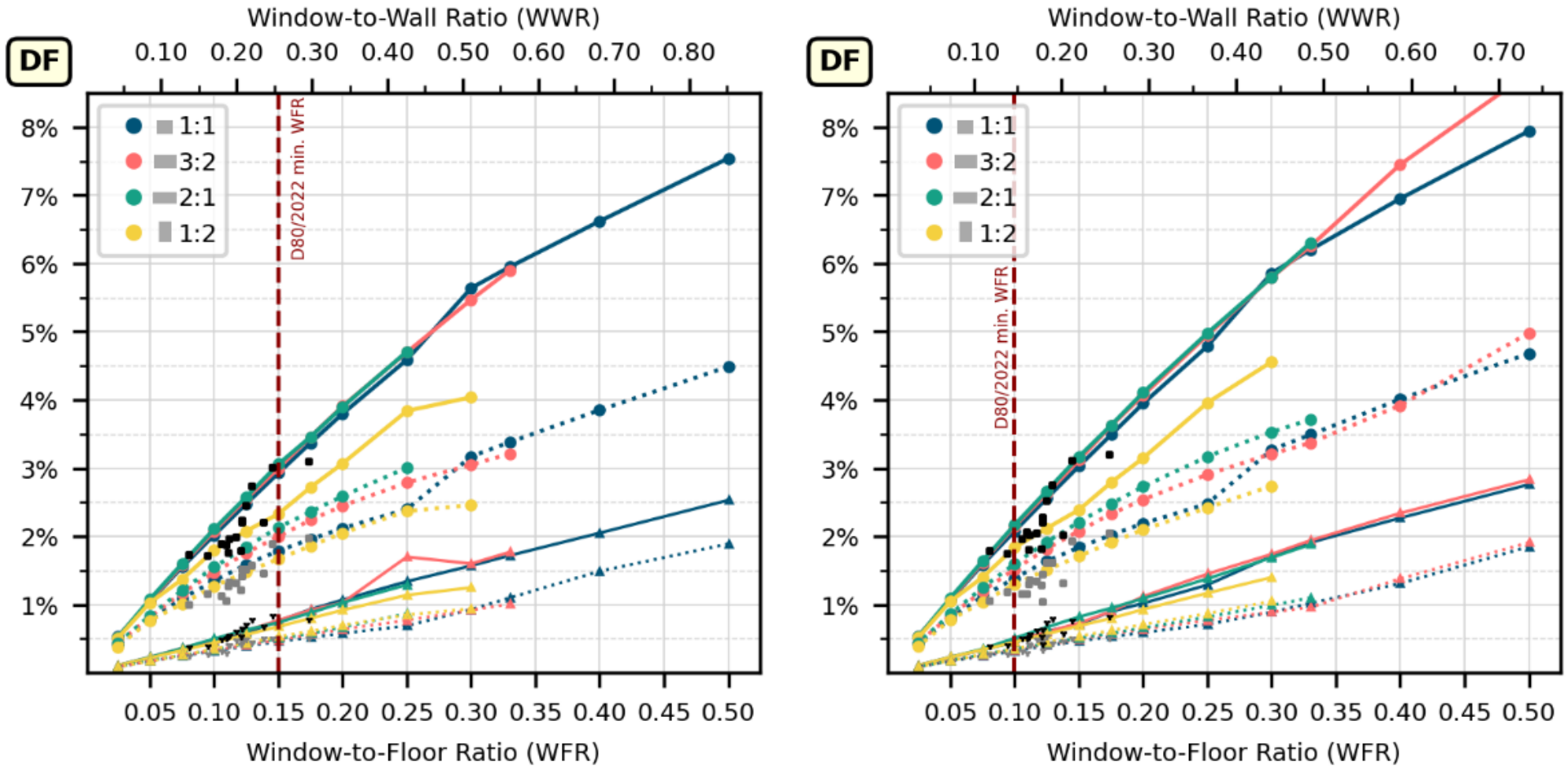

*Figure 4. Average and minimum DF for different window-to-floor ratios in BR1 (left) and BR2 (right). The solid lines with circular marks indicate ADF, while the triangular markers indicate the minimum DF. Dashed lines represent results for rooms with a 1m deep balcony.*

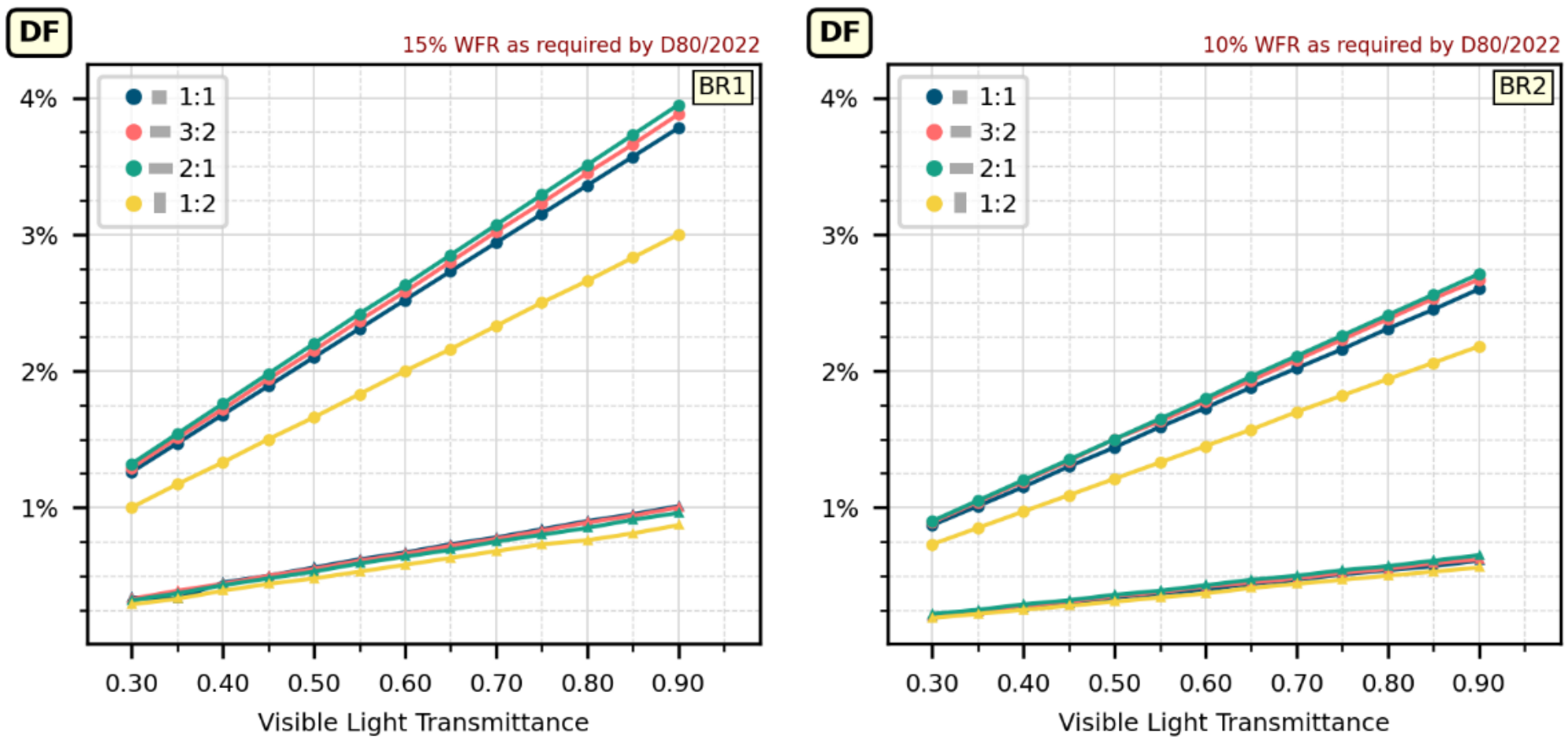

*Figure 5. Average and minimum DF for BR1 (left) and BR2 (right) with varying visible light transmittance of the insulating glazing units.*

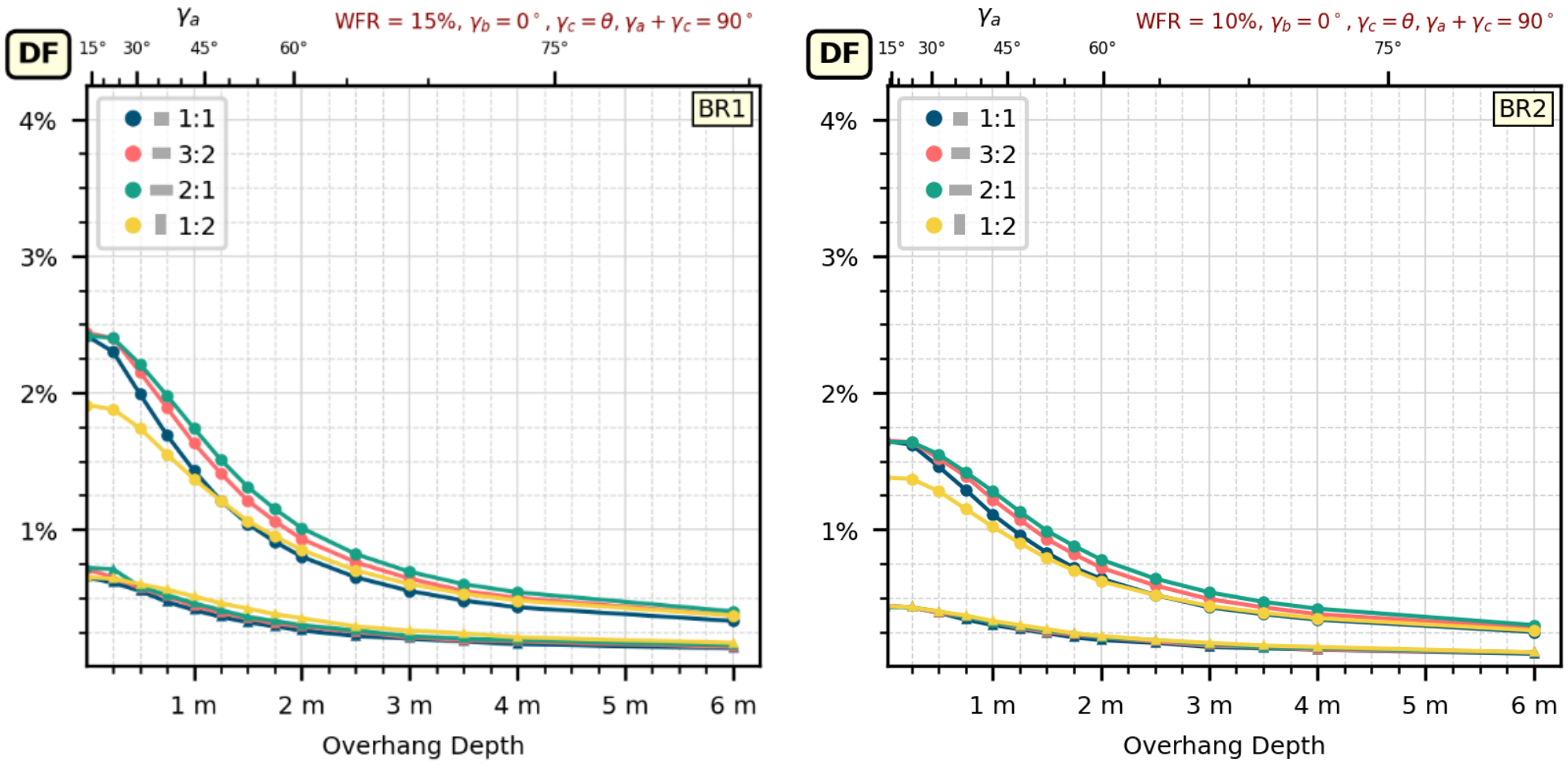

*Figure 6. Influence of overhang depth in average and minimum DF for BR1 (left) and BR2 (right).*

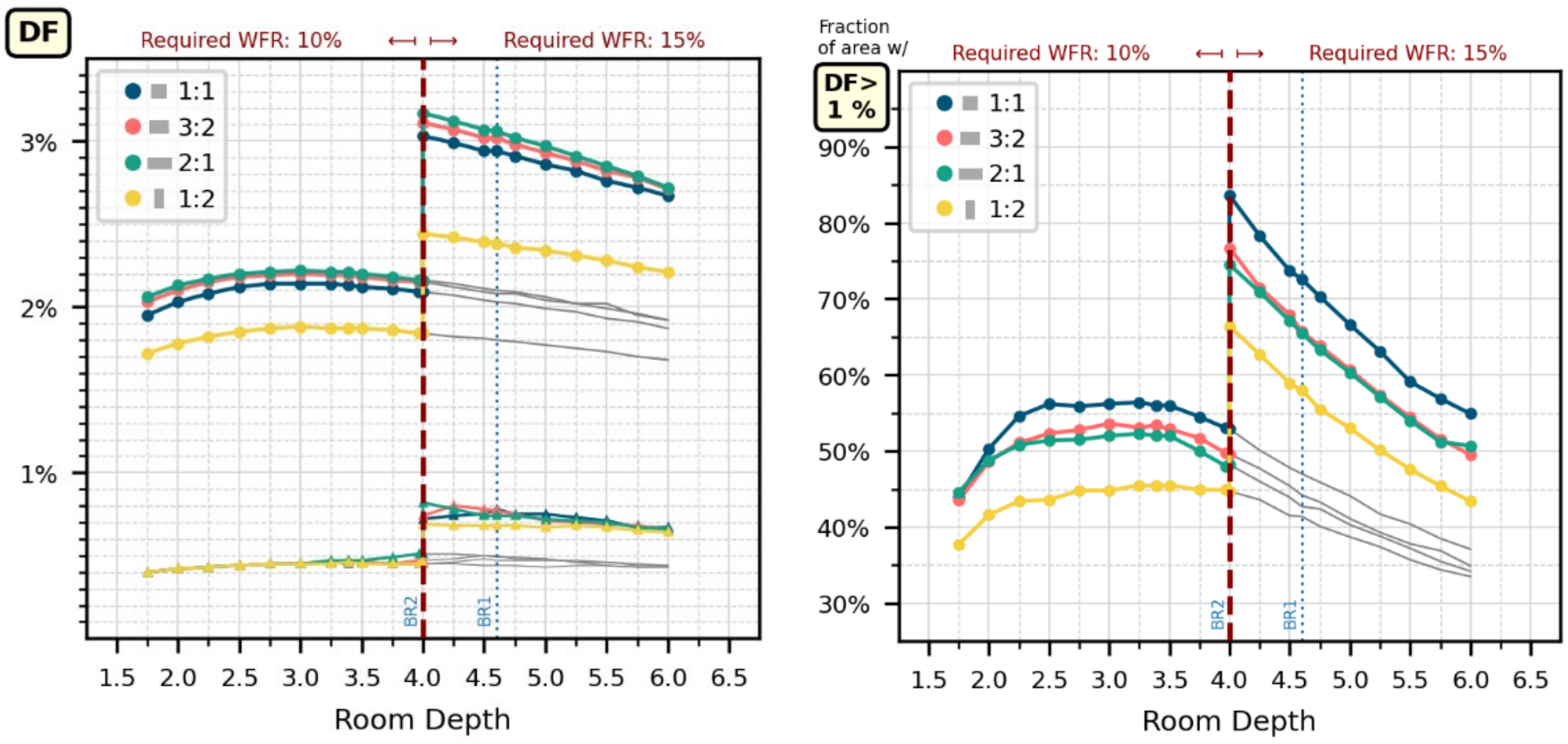

*Figure 7. Requirement curve, or translation of the requirements of daylighting in the Basque Decree to ADF (left) and to fraction of area with DF > 1% (right).*